\documentclass[twocolumn,showpacs,amsmath,amssymb]{revtex4}

\usepackage{latexsym}
\usepackage{amssymb}
\usepackage{amsfonts}
\usepackage{amsmath}
\usepackage[dvips]{graphicx}
\usepackage{bm}

\begin{document}
\title{On the time delay in binary systems}
\author{Angelo Tartaglia$^{1,2 }$, Matteo Luca Ruggiero$^{1,2 }$ and Alessandro Nagar$^{1,3 }$}
\affiliation{$^{1}$ Dipartimento di Fisica, Politecnico di Torino,
                    Corso Duca degli Abruzzi 23, 10129 Torino, Italy\\
             $^{2}$ INFN, Sezione di Torino, Via Pietro Giuria 1 Torino, Italy\\
             $^{3}$Departament d'Astronomia i Astrof\'{\i}sica, Universitat de Val\`encia,\\
                    Edifici d'Investigaci\'o, Dr.~Moliner 50, 46100 Burjassot (Val\`encia), Spain }

\date{\today}

\begin{abstract}
 The aim of this paper is to study the time delay on electromagnetic signals propagating across a
 binary stellar system. We focus on the antisymmetric gravitomagnetic contribution due to the angular
 momentum of one of the stars of the pair. Considering a pulsar as the source of the signals, the
 effect would be manifest both in the arrival times of the pulses and in the frequency shift of their
 Fourier spectra. We derive the appropriate formulas and we discuss the influence of different
 configurations on the observability of gravitomagnetic effects. We argue that the recently discovered
 PSR J0737-3039 binary system does not permit the detection of the effects because of the large size of
 the eclipsed region.
\end{abstract}


\maketitle

\section{Introduction}

\label{sec:intro}

Gravitomagnetic effects are perhaps the most elusive phenomena
predicted by General Relativity (GR). These effects are originated
by the rotation of the source of the gravitational field, which
gives rise to the presence of off-diagonal $g_{0i}$ terms in the
metric tensor. The gravitational coupling with the angular
momentum of the source is indeed much weaker than the coupling
with mass alone (gravito-electric interaction). Considering an
axisymmetric stationary configuration, we may compare the relevant
terms of the metric tensor by looking at the ratio
\begin{equation}
\varepsilon =\frac{g_{0\phi }}{g_{00}} \;,  \label{rapporto}
\end{equation}%
where a polar non-coordinated basis is assumed with unit forms
$\omega ^{0}=cdt$ and $\omega ^{\phi }=rd\phi $. Almost everywhere
in the universe a weak field approximation is acceptable, hence
\begin{align}  \label{eq:evalmetric}
g_{00}&=1-\frac{R_{S}}{r}\;, \\
g_{0\phi}&=\frac{a R_{S}}{r^{2}}\sin ^{2}\theta\;,
\end{align}
where $R_{S}=2GM/c^2$ is the Schwarzschild radius of the source, $M$ being its
mass; we have defined $a=J/(Mc)$ where $J$ is the source angular momentum.
In the equatorial plane ($\theta =\pi /2$),
Eq.~(\ref{rapporto}) reads
\begin{equation}  \label{epsilon}
\varepsilon =\frac{a R_{S}}{r\left( r-R_{S}\right) }\simeq \frac{a R_{S}}{%
r^{2} }\;.
\end{equation}
Evaluation of Eq.~(\ref{epsilon}) at the surface of the Sun
(the most favorable place in the solar system) gives $\varepsilon \sim 10^{-12}$,
thus evidencing the weakness of the gravitomagnetic versus the gravitoelectric interaction.

The smallness of $\varepsilon $ is the reason why, though having been
suggested from the very beginning of the relativistic age~\cite{lense},
the explicit verification of the existence of gravitomagnetic effects has
been extremely limited so far. Although a number of proposals for experimental tests of
gravitomagnetism have been put forward during the past decades~\cite{gravmag1,gravmag2},
the only one presently under way is the Gravity Probe B (GPB) mission~\cite{gpb}, which
is currently collecting scientific data to verify both the Lense Thirring and the
Schiff~\cite{schiff60} precessions of orbiting gyroscopes.
Other existing experimental tests of gravitomagnetism are:
\begin{itemize}
\item lunar laser ranging \cite{lunar}\;;
\item laser ranging of terrestrial artificial satellites LAGEOS
and LAGEOS II \cite{ciufolini} .
\end{itemize}
Ratio~(\ref{epsilon}) can be less unfavorable whenever $r$ is
approaching the Schwarzschild radius of the source: this can be
the case of a source of electromagnetic (e.m.) signals orbiting around a compact,
collapsed object, as it may happen in a compact binary system where (at least)
one of the stars is a pulsar.

The aim of this paper is to discuss the time delay in the
propagation of e.m. signals in GR. It is well known that the
curvature of spacetime produces a delay in the  propagation time
of light with respect to a flat environment (Shapiro delay), a
phenomenon that has been measured within the solar system
\cite{shapiro71,marte1,marte2}.

In the presence of a rotating source, a specific gravitomagnetic
contribution to the delay is also expected, which would show up as
an asymmetry in the time of flight of the signals. In
Ref.~\cite{noi} it was proposed to look for this effect in the
vicinity of the Sun. However, the magnitude of the effect is
really tiny, $\sim 10^{-10}$ s from opposite sides of the solar
disk. Other proposals pertain to the measurement of the frequency
shift induced on e.m. signals by the gravitomagnetic field of the
Sun~\cite{bertotti}.

As we pointed out above, a more favorable situation could be
expected in a compact binary system. Presently, just a few of them
are known~\cite{stairs}, but all are interesting laboratories for
testing GR.
Among the known systems, the recently discovered PSR~J0737-3039~\cite{j07}
presents a favorable configuration and is particularly appealing also
because both stars are pulsars. The data collection is
going on and maybe some interesting results can be found also with respect
to gravitomagnetic effects: for example, it has been recently argued
that both the precession of the spinning bodies and the spin
effects on the orbit could be measured in this system~\cite{Oconnell04}.

In this paper we derive the gravitational time delay
on e.m. pulses in a binary system, focusing on the gravitomagnetic contribution,
and propose how its consequences could be revealed. The corresponding frequency
shift is also briefly discussed.

The organization of this paper is as follows. In
Sec.~\ref{sec:timedelay} we describe the geometrical background
and the hypotheses assumed to calculate the time delay. In
Sec.~\ref{sec:thesystem} we review the properties of the binary
system PSR~J0737-3039, pointing out its relevance for experimental
tests of GR. In Sec.~\ref{sec:model} we apply the developed
formalism to a PSR~J0737-3039-like binary system and in
Sec.~\ref{sec:conclusion} we present our conclusions.

\section{Time delay from a binary system}
\label{sec:timedelay}

We shall refer to two objects, composing the binary system, with notation $%
\mathcal{O}_{1}$ and $\mathcal{O}_{2}$. Object $\mathcal{O}_{2}$
is supposed to be rotating (e.g. a rotating neutron star) and is
then the source of the gravitomagnetic field. The other object
$\mathcal{O}_{1}$ (e.g. the radio-pulsar) plays the role of the
source of e.m. beams. Thus, we shall focus on time-delay observed
on the e.m. signals emitted by $\mathcal{O}_{1}$ when they
experience the  field generated by $\mathcal{O}_{2}$. Furthermore,
we shall consider observers that are  far away from the source of
the gravitational field, so that they do not feel its effects,
e.g. Earth-based observers. In this case,  the coordinate time
corresponds to the proper time measured by the observers.

The derivation of the expressions relative to such a configuration
relies on some standard assumptions, listed from \emph{i}) to
\emph{vii}) in the following.

\begin{figure}[t]
\begin{center}
\includegraphics[width=75 mm,height=75 mm]{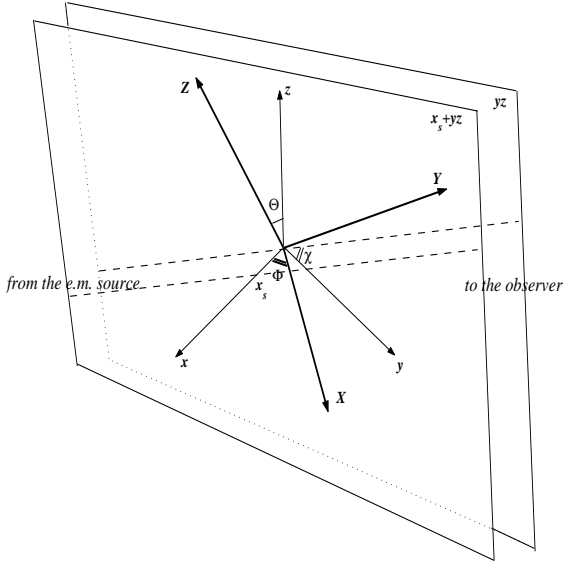}
\end{center}
\caption{Gravitomagnetic reference frame $xyz$. The origin is located on
${\cal O}_{2}$ and the $z$-axis is aligned with the direction of its
angular momentum. The Cartesian reference frame $XYZ$ is located in
${\cal O}_{2}$ and $XY$ identifies the orbital plane of the binary system.}
\label{fig:fig1}
\end{figure}

\emph{i}) We choose Cartesian coordinates, whose origin is located on the mass
which is the source of the gravitomagnetic field
(i.e., object ${\cal O}_{2}$): the $z$-axis is aligned with the direction of
the angular momentum $\overrightarrow{J}$ of the source (see Fig.~\ref{fig:fig1}),
the $x$-axis is orthogonal to the line of sight from Earth and the $y$-axis
is orthogonal to both; we shall refer to $xyz$ as to the ``gravitomagnetic''
reference frame. Consequently, the line element reads
\begin{align}
ds^{2}& =g_{00}c^{2}dt^{2}+g_{xx}dx^{2}+g_{yy}dy^{2}  \notag \\
& +g_{zz}dz^{2}+2g_{0x}cdxdt+2g_{0y}cdydt \;.
\label{linea}
\end{align}

\emph{ii}) The gravitational field is in any case weak enough to admit the
approximation ($r=\sqrt{x^{2}+y^{2}+z^{2}}$)
\begin{eqnarray}
g_{00} &=& 1-\frac{R_{S}}{r}\;, \label{metrica11}      \\
g_{xx} &=& g_{yy}=g_{zz}=-1-\frac{R_{S}}{r}\;,  \label{metrica}
\end{eqnarray}
\begin{eqnarray}
g_{0x} &=& -\frac{aR_{S}y}{r^{3}}\;, \label{metrica12} \\
g_{0y} &=& \frac{aR_{S}x}{r^{3}}\;. \label{metrica13}
\end{eqnarray}

\emph{iii}) The trajectory of light rays is assumed to be a
straight line. Actually, there is a bending, whose effects are
usually assumed to be negligible~\cite{straumann84}, \footnote{In
fact: {\emph a}) The increase in the time of flight purely due to
the geometric lengthening of light rays because of the bending
induced by curvature is relevant only within the binary system and
is quadratic in the deviation angle $\alpha$. The deviation angle
is in turn proportional to the ratio between the mass of one of
the stars and the impact parameter, which gives in practice a
quadratic factor in the order of $10^{-9}$; {\emph b}) The
relevant quantity is the interval between two consecutive pulses
from the source. The influence of the bending would depend on the
change of the deviation angle during one period of the signal. In
practice, this change introduces another factor $\sim 10^{-4}$ or
less. The final impact is then less then one part in $10^{13}$;
{\emph c}) The effect due to the mass is symmetric with respect to
conjunction, then it is irrelevant for gravitomagnetic
contributions; {\emph d}) There is an anisotropy in the deviation
angle too, but this effect is smaller than the very deviation.}.

\emph{iv}) The center of mass of the system is at rest with respect to the
observer on Earth.

\emph{v}) The orbit of the source of the signals (i.e., object $\mathcal{O}_1
$) around the center of mass of the system is circular \footnote{This condition
is not really necessary, but simplifies the description of the system.}.

\emph{vi}) The size of the binary system is much smaller than the distance
from the observer on Earth.

\emph{vii}) As a result of the proper motion of object $\mathcal{O}_2$, the origin
of the reference frame is moving with respect to the center of mass of the system
and then with respect to the observer. However, we shall assume that this motion is
slow enough not to appreciably change the expression~(\ref{metrica}) of the metric.
In practice, from the viewpoint of the observer, the propagation of light through
the system is described as a series of static snapshots.

Under these conditions, we identify the position of the source of the signals
with the space coordinates $\left( x_{\rm s},y_{\rm s}, z_{\rm s}\right) $
and that of the observer with $\left( x_{\rm s}, y_{\rm obs},z_{\rm obs}\right)$.
According to hypothesis \emph{iii}), the trajectory of the e.m. beam is a
straight line
\begin{eqnarray}
x &=&x_{\rm s}\;,   \\
z &=&z_{\rm s}+\left( y-y_{\rm s}\right) \tan \chi \;.
\label{retta}
\end{eqnarray}
Each beam, represented by a dashed line in Fig.~\ref{fig:fig1},
lies in a plane parallel to $yz$, thus
\begin{equation}
0=g_{00}c^{2}dt^{2}+g_{yy}h^{2}dy^{2}+2g_{0y}cdydt\;,  \label{line1}
\end{equation}%
and in the components of the metric tensor
(\ref{metrica11})-(\ref{metrica13}) we can replace
$r^{2}=h^{2}y^{2}+2ky+w^{2}$, with
\begin{eqnarray}
h &=&\sqrt{1+\tan ^{2}\chi }\;, \\
k &=&\left( z_{\mathrm{s}}-y_{\mathrm{s}}\tan \chi \right) \tan \chi \;, \\
w &=&\sqrt{\left( z_{\mathrm{s}}-y_{\mathrm{s}}\tan \chi \right) ^{2}+x_{%
\mathrm{s}}^{2}}\;.
\end{eqnarray}
According to the standard approach to the time delay problem~\cite{straumann84},
we solve Eq.~(\ref{line1}) for $dt/dy$, then the result is integrated along
the trajectory of the ray; i.e.,
\begin{align}
t_{\mathrm{flight}}& =\dfrac{1}{c}\int_{y_{\mathrm{s}}}^{y_{\mathrm{obs}%
}}dy\;\left( r-R_{S}\right) ^{-1}  \notag \\
& \times \Bigg\{-\dfrac{aR_{S}x_{\mathrm{s}}}{r^{2}}+\sqrt{\frac{%
a^{2}R_{S}^{2}x_{\mathrm{s}}^{2}}{r^{4}}+\left( r^{2}-R_{S}^{2}\right) h^{2}}%
\Bigg\}\;.  \label{tof}
\end{align}%
When the propagation is ``on the left'' of
the oriented projection of $\overrightarrow{J}$ in the sky, with respect to
the observer ($x_{\mathrm{s}}>0$, see Fig.~\ref{fig:fig1}), the first term
in the parenthesis is negative; on the opposite, when the propagation is on
the right ($x_{\mathrm{s}}<0$), the sign is positive.

For the weak field condition of~\emph{ii}), Eq.~(\ref{tof}) can be further
expanded in powers of $R_{S}$ and $a$, up to their product, i.e., up to second
order. Second order is necessary to describe the gravitomagnetic interaction;
in addition, it should be noticed that, for collapsed objects (as for a star
like the Sun) it is reasonably $R_{S}\sim a$, so that the second order term
in $R_{S}$ cannot be simply neglected. By performing such an expansion, the
integral of Eq.~(\ref{tof}) is divided into four terms, which may be further
grouped into three contributions to the total time of flight as
\begin{equation}
\delta t\equiv t_{\mathrm{flight}}=t_{0}+t_{M}+t_{J}\;,  \label{eq:tof1}
\end{equation}
where
\begin{equation}
t_{0}=\frac{1}{c}\int_{y_{\mathrm{s}}}^{y_{\mathrm{obs}}}hdy=\frac{h}{c}%
\left( y_{\mathrm{obs}}-y_{\mathrm{s}}\right)   \label{geometrico}
\end{equation}%
represents the pure geometric term,
\begin{align}
t_{M}& =\frac{h}{c}R_{S}\int_{y_{\mathrm{s}}}^{y_{\mathrm{obs}}}\left( \frac{%
1}{r}+\frac{R_{S}}{2r^{2}}\right) dy  \notag \\
& =\frac{R_{S}}{c}\ln \frac{h^{2}y_{\mathrm{obs}}+k+hr_{\mathrm{obs}}}{%
h^{2}y_{\mathrm{s}}+k+hr_{\mathrm{s}}}  \label{massa} \\
& +\frac{1}{2c}\frac{hR_{S}^{2}}{\sqrt{h^{2}w^{2}-k^{2}}}\left. \allowbreak
\arctan \left( 2\frac{k+h^{2}y^{2}}{\sqrt{h^{2}w^{2}-k^{2}}}\right)
\right\vert _{y_{\mathrm{s}}}^{y_{\mathrm{obs}}}  \notag
\end{align}%
is the mass delay up to second order, where we have defined $r_{\mathrm{obs}%
}\equiv r(y_{\mathrm{obs}})$ and $r_{\mathrm{s}}\equiv r(y_{\mathrm{s}})$; the
contribution due to the angular momentum of the source reads
\begin{align}
\label{angolare}
t_{J}=-\frac{x_{\mathrm{s}}}{c}\int_{y_{\mathrm{s}}}^{y_{\mathrm{obs}}}
\frac{aR_{S}}{r^{3}}dy
=\frac{x_{\mathrm{s}}}{c}\frac{aR_{S}}{k^{2}-h^{2}w^{2}}%
\left. \frac{h^{2}y+k}{r}\allowbreak \right\vert _{y_{\mathrm{s}}}^{y_{%
\mathrm{obs}}}\;.
\end{align}
The quantity $x_{\mathrm{s}}$ changes sign if the e.m.
source is on the left or on the right of the rotating body with respect to
the observer; as a result, $t_{J}$ can have opposite signs on opposite sides
accordingly.

Focusing on the geometry peculiar of a binary system, one is
expecting $y_{\mathrm{obs}}$ to be the sum of a time-independent
part, $y_{0} $, corresponding to the distance from the center of
mass of the system to the observer, and a time-dependent part, a
contribution oscillating in time due to the orbital motion of
object $\mathcal{O}_{2}$;  condition $v$) implies that this orbit
too is a circumference of radius $R_{2}$. The amplitude of the
oscillation is of the order of magnitude of the size of the binary
system, just as $k$ and $w$. Since for condition \emph{vi}) the
size of the system is much smaller than the distance from Earth,
Eqs.~(\ref{massa}) and (\ref{angolare}) are simplified as
\begin{align}
t_{M}& \simeq \frac{R_{S}}{c}\ln \frac{2y_{0}h^{2}}{h^{2}y_{\mathrm{s}%
}+k+hr_{\mathrm{s}}}  \notag  \label{massa1} \\
& +\frac{1}{2c}\frac{hR_{S}^{2}}{\sqrt{h^{2}w^{2}-k^{2}}}  \notag \\
& \times \left\{ \arctan \frac{2h^{2}y_{0}^{2}}{\sqrt{h^{2}w^{2}-k^{2}}}%
-\arctan \frac{2(k+h^{2}y_{\mathrm{s}}^{2})}{\sqrt{h^{2}w^{2}-k^{2}}}%
\right\},  \\
t_{J}& \simeq
\frac{x_{\mathrm{s}}}{c}\frac{aR_{S}}{k^{2}-h^{2}w^{2}}\left(
h\allowbreak -\frac{h^{2}y_{\mathrm{s}}+k}{r_{\mathrm{s}}}\right)\;,
\label{angolare1}
\end{align}
where we have set
$h^{2}y_{\mathrm{s}}^{2}+2ky_{\mathrm{s}}+w^{2}=r_{\mathrm{s}}^{2}=R^{2}$;
$R$ is the distance between the two stars in the pair, which is constant
according to hypothesis \emph{v}). By restoring an explicit notation,
we have
\begin{eqnarray}
\label{massabuono}
t_{M} &=&t_{M_1}+t_{M_2}\simeq \frac{R_{S}}{c}\ln \frac{2y_{0}}{\allowbreak
\left( y_{\mathrm{s}}\cos \chi +z_{\rm s}\sin \chi +R\right) \cos \chi }\nonumber\\
&+&\frac{1}{2c}\frac{R_{S}^{2}}{\sqrt{R^{2}-\left( y_{\mathrm{s}}\cos \chi
+z_{\mathrm{s}}\sin \chi \right) ^{2}}} \\
&\times&\left( \allowbreak \frac{\pi }{2}-\arctan \frac{2\left( \allowbreak
y_{\mathrm{s}}\cos \chi +z_{\mathrm{s}}\sin \chi \right) }{\sqrt{%
R^{2}-\left( y_{\mathrm{s}}\cos \chi +z_{\mathrm{s}}\sin \chi \right) ^{2}}}%
\right) \;,  \nonumber\\
t_{J}&\simeq& -\frac{x_{1}}{c}\frac{aR_{S}}{R}\frac{\cos \chi }{R+\allowbreak
y_{\rm s}\cos \chi +z_{\rm s}\sin \chi } \;.   \label{angbuono}
\end{eqnarray}

\subsection{The time dependent part}
\label{ssec:tdeppart}
In Eqs.~(\ref{geometrico}), (\ref{massabuono}), and (\ref{angbuono}) we are
interested in the time dependent part, which is implicit in $x_{\mathrm{s}}$%
, $y_{\rm s}$ and $z_{\rm s}$, since the position of the e.m.
source at each successive ``snapshot'' is different because of the orbital
motion of $\mathcal{O}_{1}$.
By assumption \emph{v}), the orbit of $\mathcal{O}_{1}$ is
circular. Let us start by expressing the position of
$\mathcal{O}_{1}$ with respect to another reference frame (called
$XYZ$, see Fig.~\ref{fig:fig1}) centered in ${\cal O}_{2}$ such
that $X=R\cos \omega t$ and $Y=R\sin \omega t$, where $\omega $ is
the orbital angular velocity of the pair and the $X$ axis is
identified by the intersection between the orbital plane of the
system and the gravitomagnetic equatorial plane of ${\cal O}_{2}$.
We call $\Phi $ the angle between the $X$-axis and the
$x$-axis; $\Theta$ identifies the tilt angle between the axis of
the orbit and the angular momentum $\overrightarrow{J}$ of
$\mathcal{O}_{2}$. The ``gravitomagnetic'' coordinates of
$\mathcal{O}_{1}$ expressed with respect to the $xyz$ frame read
\begin{align}
\xi _{\rm s}&   =\cos \Phi \cos \psi -\cos \Theta \sin \Phi \sin \psi \;,
\notag  \label{orbita1} \\
\eta _{\rm s}&  =\sin \Phi \cos \psi +\cos \Theta \cos \Phi \sin \psi \;,\\
\zeta _{\rm s}& =-\sin \Theta \sin \psi \;,  \notag
\end{align}
where, for convenience, we have introduced the reduced coordinates
$\xi _{\rm s}=x_{\mathrm{s}}/R$, $\eta _{\mathrm{s}}=y_{\mathrm{s}}/R$, $\zeta
_{\mathrm{s}}=z_{\mathrm{s}}/R$, and we have defined $\psi =\varphi+\alpha$:
$\varphi=\omega t$ is the orbital phase and
$\alpha =\arctan \left( \cot \Phi /\cos \Theta \right)$.
We can measure times from the configuration
$x_{s}=0$, $y_{s}>0$ (conjunction).

From Eq.~(\ref{geometrico}), the time-dependent contribution $t^*_0$
in flat spacetime reads (hereafter, starred quantities refer to the time-dependent contributions
to $t_{\rm flight}$)
\begin{equation}
\frac{ct_{0}^{\ast }}{r-R}\cos \chi =\sin \Phi \cos \psi +\cos \Theta \cos
\Phi \sin \psi\;,  \label{geo}
\end{equation}
that is, a harmonic oscillation whose amplitude corresponds to the time
the beam takes to cross the system.

The mass-dependent term is more involved, since it is composed by a first
order and a second order term. Both of them can be factorized into an
``amplitude'', containing the size of the system and the Schwarzschild
radius of $\mathcal{O}_{2}$, and a pure geometrical part. It is given by
the sum $t_{M}^*=t_{M_{1}}^*+t_{M_{2}}^*$, with
\begin{eqnarray}
\frac{ct_{M_{1}}^{\ast }}{R_{S}} &=&-\ln \left( \eta _{\mathrm{s}}\cos \chi
+\zeta _{\mathrm{s}}\sin \chi +1\right) \;, \\
\frac{ct_{M_{2}}^{\ast }}{R_{S}^{2}} &=&\frac{1}{2R\sqrt{1-\left( \eta _{%
\mathrm{s}}\cos \chi +\zeta _{\mathrm{s}}\sin \chi \right) ^{2}}}
\label{dimassa} \\
&\times &\left( \allowbreak \frac{\pi }{2}-\arctan \frac{2\left( \eta _{%
\mathrm{s}}\cos \chi +\zeta _{\mathrm{s}}\sin \chi \right) }{\sqrt{1-\left(
\eta _{s}\cos \chi +\zeta _{\mathrm{s}}\sin \chi \right) ^{2}}}\right) \;.
\notag
\end{eqnarray}%
Eventually, from Eq.~(\ref{angolare}), the contribution to the time delay
due to the angular momentum of $\mathcal{O}_{2}$ reads
\begin{equation}
\frac{ct_{J}}{aR_{S}}\frac{R}{\cos \chi }\simeq -\frac{\xi _{\mathrm{s}}}{%
\left( 1+\eta _{\mathrm{s}}\cos \chi +\zeta _{\mathrm{s}}\sin \chi \right) }%
\;.  \label{eq:tj11}
\end{equation}
These time-dependent parts of $t_{\rm flight}$ would be visible in the sequence
of the arrival times of the pulses from the source $\mathcal{O}_{1}$.

\section{The binary system PSR J0737-3039}
\label{sec:thesystem}
\begin{table}[t]
\caption{\label{label:table1}A toy model for system PSR
J0737-3039. We choose $\omega\sim 7\times 10^{-4}$ s$^{-1}$ and
$R=2R_2\simeq 10^9$ m. From left to right, the columns contain
which star of the pair is the source of the gravitomagnetic field,
the Schwarzschild radius $R_S$ of ${\cal O}_2$ and the angles that
identify the geometrical configuration of the system.}
\begin{ruledtabular}
\medskip
\begin{tabular}{cccccc}
${\cal O}_1$ & ${\cal O}_2$  & $R_S$ (m)  & $\chi$ & $\Theta $ & $\Phi$ \\
\hline
${\rm A}$  &  B      &  $1.6\times 10^3$  & $0^{\circ}$  &  $0^{\circ}$ & $0^{\circ}$ \\
${\rm B}$  &  A       & $1.7\times 10^3$  & $50^{\circ}$ & $50^{\circ}$ & $0^{\circ}$ \\
\end{tabular}
\end{ruledtabular}
\end{table}
The recently discovered binary system PSR J0737-3039 has proved to
be an important laboratory for testing relativistic
theories and, in principle, it could be useful also for measuring
gravitomagnetic effects on time delay. Let us then briefly
review its most important physical features~\cite{j07}.
The two pulsars, J0737-3039A and J0737-3039B (hereafter simply A and B),
have periods $P_{A}=23$ ms and $P_{B}=2.8$ s;
they revolve about each other in a 2.4-hr orbit of
significant eccentricity (0.088); the separation of the two
objects is typically $9\cdot 10^{5}\ $km. The orbital plane is
viewed nearly edge-on from the Earth, with an inclination angle of
$i=87^{\circ }\pm 3^{\circ }$. It has been
possible to detect a huge rate of periastron advance, $\dot{\omega}%
=16.88^{\circ }$ yr$^{-1}$, which is about four times the one of
PSR 1913+16 \cite{taylor79}. If this effect is
entirely due to GR, from the observations carried
out so far it has been possible to establish that
$M_{A}=1.337(5)M_{\odot }$ and $M_{B}=1.250(5)M_{\odot }$. In
addition, due to the collision of A's wind with B's magnetosphere
it seems very likely that the spin axis of B is aligned with the
orbital angular momentum of the system~\cite{demorest04,arons04}.
On the other hand, observations show that A is almost an aligned
rotator (angle between A's magnetic and rotation axes $\sim
5^{\circ }$~\cite{demorest04}), but with
its spin axis substantially misaligned with the orbital angular momentum
by $\sim 50^{\circ }$. In addition, the system has the important feature that,
for $27$ s, A is eclipsed by B's magnetosphere. Such duration of the eclipse
was used in Ref.~\cite{kaspi04} to place a limit of $18.6\times 10^3$ km on the size
of the eclipsed region. This region is much bigger than the
expected  physical linear dimensions of an actual neutron star ($\sim 10$ km):
the typical features of the observed signals seem to suggest that the eclipse
is due to the absorption of the radio emission from A by a
magnetosheath surrounding B's magnetosphere~\cite{maclaughlin04,arons04,lyutikov04}.

Because of (i) the alignment of the orbital plane with the line of
sight, (ii) the fact that B eclipses A and (iii) the spin axis of
B is probably perpendicular to the orbital plane, the
configuration of the system could be favorable for studying the
gravitomagnetic effects on signals propagation.

\section{Application to a PSR J0737-3039-like model}
\label{sec:model}
\subsection{The time delay}
Let us specify the simple model outlined above using
values of the parameters similar to those of the actual
PSR J0737-3039: star B is acting as the source of gravitomagnetism
(${\cal O}_2\equiv B$ and ${\cal O}_1\equiv A$, see Table \ref{label:table1}),
$R \sim 10^{9}$ m and $ \omega \sim 7\times 10^{-4}\text{s}^{-1}$.
The angular momentum of star B is aligned with the orbital angular
momentum of the system ($\Theta=0$) and we choose the most favorable
configuration  with $\Phi=0$ (i.e. $\alpha=\pi/2$). The e.m beams are
thus propagating in the orbital plane ($\chi=0$).
Supposing that the progenitor star was only a little bigger than
the Sun, and that most of the angular momentum was
preserved during the collapse, we can assume  $a\sim 10^{3}\text{m}$.
With these hypotheses, Eqs. (\ref{geo})-(\ref{eq:tj11}) read
\begin{eqnarray}
t_{0}^{\ast } & = & \mathcal{A}^{\rm B}_0 \cos\varphi  \label{deltatzerob}, \\
t_{M_1}^{\ast } & = & \mathcal{A}^{\rm B}_{M_1} \ln(1+\cos\varphi)\label{tiemme1b}, \\
t_{M_2}^{\ast } & = &\mathcal{A}^{\rm B}_{M_2} \frac{1}{\left\vert
\sin \varphi\right\vert }\left( \allowbreak \frac{\pi }{2}-\arctan \frac{2\cos \varphi%
}{\left\vert \sin \varphi\right\vert }\right) \label{tiemme2b}, \\
t_{J} & = & \mathcal{A}^{\rm B}_{J} \frac{\sin \varphi}{1+\cos\varphi}\label{tjb1}
\end{eqnarray}
where the numerical coefficients ${\cal A}^{{\cal O}_2}$ with ${\cal O}_2\equiv {\rm B}$
give the order of magnitude of the effect (see Table~\ref{label:table2}).
\begin{table}[t]
\caption{\label{label:table2}The first two rows show the order of
magnitude of the contributions to the time-dependent part of the time of
flight of e.m. signals emitted by object ${\cal O}_1$ and propagating in
the gravitational field generated by ${\cal O}_2$.
The bottom row contains the order of magnitude  of the contributions
to the (relative) frequency shift on the signals emitted by star A.}
\begin{ruledtabular}
\medskip
\begin{tabular}{cccccc}
${\cal O}_1$ & ${\cal O}_2$  & $\mathcal{A}^{{\cal O}_2}_0$ (s) & $\mathcal{A}^{{\cal O}_2}_{M_1}$ (s) & $\mathcal{A}^{{\cal O}_2}_{M_2}$ (s)  & $\mathcal{A}^{{\cal O}_2}_{J}$ (s) \\
\hline\hline
${\rm A}$  &  B      & $-3$  &  $-5\times 10^{-6}$      &  $4\times 10^{-12}$   & $6\times 10^{-13}$\\
${\rm B}$  &  A       & $-4.7$  &  $-5.6\times 10^{-6}$ & $4.8\times 10^{-12}$  & $-3.6\times 10^{-12}$  \vspace{.1cm} \\
\hline\hline
  & & ${\cal D}^{{\cal O}_2}_{0}$  & ${\cal D}^{{\cal O}_2}_{M_1}$  & ${\cal D}^{{\cal O}_2}_{J_1}$
& ${\cal D}^{{\cal O}_2}_{J_2}$ \\
\hline\hline
${\rm A}$  &  B      & $-21\times 10^{-4}$  &  $-3.7\times 10^{-9}$      &  $-4\times 10^{-15}$  & $4\times 10^{-15}$   \\
\end{tabular}
\end{ruledtabular}
\end{table}
The $\varphi$-dependent parts in Eqs.~(\ref{tiemme1b}) and (\ref{tjb1}) become
bigger and bigger close to the conjunction position ($\varphi=\pi$; i.e., when the
impact parameter is zero), but this divergence has no physical meaning because
the actual compact objects have finite dimensions and the beam can not pass through
the center of B.

\begin{figure}[t]
\begin{center}
\includegraphics[width=8cm,height=9cm]{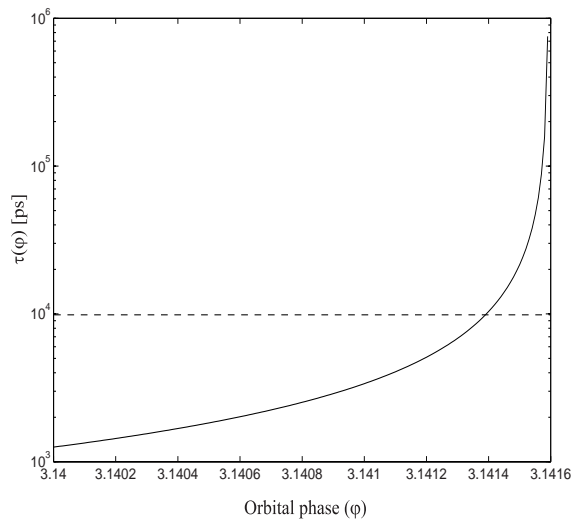}
\end{center}
\caption{The function $\tau\left( \varphi \right)$ (measured in
picoseconds) versus the orbital phase of star A in proximity of
the occultation position $\varphi=\pi$. The horizontal dashed line
corresponds to the detectability threshold of $10^{-8}$ s.}
\label{fig:fig2}
\end{figure}

Let us suppose that it is possible to identify conjunction ($\varphi=\pi$) and
opposition ($\varphi=2\pi$) points in the sequence of arriving pulses. Since the geometric
and mass terms are symmetric with respect to conjunction and opposition,
whereas $t_{J}$ is antisymmetric, for $0\leq \varphi \leq \pi $
we have
\begin{align}
\label{antisim1}
\tau\left( \varphi \right)&=\delta t^{\ast}
\left( \varphi \right) -\delta t^{\ast} \left( 2\pi -\varphi\right)=2t_{J},
\end{align}
i.e., in seconds
\begin{align}
\label{antisim2} \tau\left( \varphi \right)&\simeq 10^{-12}%
\frac{\sin \varphi }{1+\cos \varphi } \;.
\end{align}
The function $\tau(\varphi)$ is shown in Fig.~\ref{fig:fig2}
(solid line) close to the conjunction position.
If we suppose that the threshold for detecting the change in the
arrival rate of the signals is for instance at $10^{-8}$ s, then
only the part of the graph above the horizontal dashed line is useful.
This means that the access to the interesting region would be possible
only if the minimum impact parameter was smaller than $\sim 1.8\times 10^{2}$
km.
Since the typical radius of a neutron star is $\sim 10$ km,
we can expect the appropriate conditions to be satisfied in a double pulsar
binary system. However, this is not the case of PSR J0737-3039: in fact, in
this system the minimum impact parameter is not given by the radius of object B,
but rather by the size of the opaque area of the magnetosheath surrounding
B itself, that is $1.8\times 10^{4}$ km; i.e. two order of magnitude bigger
than the detectability threshold.

The same kind of analysis can be performed by exchanging A with B;
i.e., ${\cal O}_2\equiv {\rm A}$ and ${\cal O}_1\equiv {\rm B}$,
although in the actual system the B pulses are extremely weak and aleatory.
In this case, we must choose $\Theta=50^{\circ}$ and $\chi=50^{\circ}$
(see Table~\ref{label:table1}) since only the e.m. beams propagating
in the orbital plane can be seen by the observer on Earth. In this case
we have
\begin{eqnarray}
t_{0}^{\ast } & = & \mathcal{A}^{\rm A}_0 \cos \varphi   \label{eq:f0a}\;, \\
t_{M_1}^{\ast } & = & \mathcal{A}^{\rm A}_{M_1} \ln  \left( 1-n\cos \varphi \right)  \label{eq:fm1a}\;,
\end{eqnarray}
\begin{eqnarray}
t_{M_2}^{\ast } & = &\mathcal{A}^{\rm A}_{M_2} \frac{1}{\sqrt{1-\left( n\cos
\varphi \right) ^{2}}}\nonumber\\
&\times&\left( \allowbreak \frac{\pi }{2}-\arctan \frac{%
m\cos \varphi }{\sqrt{1-\left( n\cos \varphi \right)
^{2}}}\right)\;, \label{eq:fm2a}  \\
t_{J} & = & \mathcal{A}^{\rm A}_{J} \frac{\sin \varphi }{1+n\cos\varphi } \label{eq:fja}\;,
\label{tja}
\end{eqnarray}
where  $n=0.17$ and  $m=0.34$, from the evaluation of Eqs.~(\ref{geo})-(\ref{eq:tj11})
employing the parameters of Table \ref{label:table1}. In this case, we obtain
\begin{equation}
\tau\left( \varphi \right) \simeq -7.2\times
10^{-12}\frac{\sin \varphi }{1+n\cos \varphi }\;. \label{antisim3}
\end{equation}
Since $|n|<1$, the denominator never vanishes, so that
the amount of the effect is at most $\sim 10^{-11}$ s.
This configuration is thence less favorable than the previous one,
because a lower detectability threshold is required.

\subsection{The frequency shift}
\label{ssec:shift}

Let us consider the Fourier spectrum of an e.m. beam propagating
into a gravitational field: the time-delay effect we have discussed
so far corresponds, in the frequency domain, to a frequency shift of
each harmonic component of the signal. Since the period $T$ of each harmonic
is much shorter than all the other characteristic time-scales of the system,
we can write its relative change in period, hence in the frequency $\nu $, as
\begin{equation}
\label{shift00}
\frac{\,\delta \nu }{\nu }=-\frac{\,\delta T}{T}=
-\left( \dot{t}_{0}+\dot{t}_{M}+\dot{t}_{J}\right) \;,
\end{equation}
where the overdot stands for derivative with respect to the coordinate time.
For each contribution, we have
\begin{eqnarray}
\left. \frac{\delta \nu }{\nu }\right\vert _{0}\;\;\, &=&\omega \frac{h}{c}%
\left( R_{2}-R\right) \sin \varphi\;,  \label{shift0} \\
\left. \frac{\delta \nu }{\nu }\right\vert _{M_{1}} &=&\frac{R_{S}}{c}\frac{%
\dot{\eta}_{\mathrm{s}}\cos \chi +\dot{\zeta}_{\mathrm{s}}\sin
\chi }{1+\eta
_{\mathrm{s}}\cos \chi +\zeta _{\mathrm{s}}\sin \chi }\;,  \label{shift1} \\
\left. \frac{\delta \nu }{\nu }\right\vert _{M_{2}} &=&
-\left(\dot{y}_{\rm s}\partial_{y_{\rm s}}+\dot{z}_{\rm s}\partial_{z_{\rm s}}\right)t_{M_2}  \\
\left. \frac{\delta \nu }{\nu }\right\vert _{J}\;\; &=& \frac{aR_S}{cR}\bigg\{
\frac{\dot{\xi}_{\rm s}\cos\chi }{1+\eta _{\rm s}\cos \chi +\zeta _{\rm s}\sin\chi }\nonumber \\
& & \qquad- \dfrac{\xi _{\rm s}\left(\dot{\eta}_{\mathrm{s}}
\cos ^{2}\chi + \dot{\zeta}_{\rm s}\sin ^{2}\chi \right)}
{\left( 1+\eta_{\rm s}\cos\chi +\zeta _{\rm s}\sin \chi \right)^{2}}\bigg\}\;,  \label{shift2}
\end{eqnarray}%
where $\dot{\xi}_{\rm s}$, $\dot{\eta}_{\rm s}$ and $\dot{\zeta}_{\rm s}$
are obtained from Eqs.~(\ref{orbita1}).
Here it becomes clear that the relative frequency shift is a complicated function of time
that reads out as a periodic modification in the frequency spectrum of the signal.

Applying the above equations to our PSR J0737-3039-like system, we just
present results for pulses emitted by star A. In this case, we have
\begin{align}
\label{zero}
\left. \frac{\delta \nu }{\nu }\right\vert _{0}&={\cal D}^{\rm B}_{0} \sin \varphi\;,\\
\left.\frac{\delta \nu }{\nu }\right\vert _{M} \!\!&={\cal D}^{\rm B}_{M_1}\frac{%
\sin \varphi }{1+\cos \varphi }-\dot{y}_{\rm s}\partial_{y_{\rm s}}t_{M_2} \label{uno}\;, \\
\left. \frac{\delta \nu }{\nu }\right\vert _{J}&={\cal D}^{\rm B}_{J_1}\frac{%
\cos \varphi }{1+\cos \varphi }
+ {\cal D}^{\rm B}_{J_2}\frac{\sin \varphi }{\left( 1+\cos\varphi\right) ^{2}}\;,
\label{due}
\end{align}
where the coefficients ${\cal D}^{{\cal O}_2}$, with ${\cal O}_2\equiv{\rm B}$,
give the order of magnitude of the effect and are listed in Table~\ref{label:table2}.

All the contributions (including $\left. \delta \nu /\nu %
\right\vert _{M_2}$ that is then needless to make explicit) are odd in
$\varphi $ except for the one proportional to ${\cal D}^{\rm B}_{J_1}$,
which is even. Summing shifts symmetric with respect to the opposition
point we obtain
\begin{align}
\frac{\delta \nu }{\nu }\left( \varphi \right) +\frac{\delta \nu
}{\nu }\left( 2\pi -\varphi \right)=-8\times 10^{-15}\frac{\cos \varphi }{1+\cos
\varphi } \label{antisimmfreq}\;.
\end{align}

\section{Conclusions}\label{sec:conclusion}

In this paper, we have discussed the effects of the gravitational
interaction on the time delay  of electromagnetic signals coming from a
binary system composed by a radio-pulsar and another compact
object. In particular, we have evidenced that the behavior of the
gravitomagnetic contribution, near the occultation of the
radio-pulsar by its companion, is antisymmetric while the
geometric and mass contribution are symmetric, thus suggesting a
possible way for decoupling the effects.

The recently discovered binary pulsar system PSR J0737-3039,
lends itself for the study of the time delay, because
(i) the orbital plane almost contains the line of sight, (ii) the star
B eclipses A and (iii) the spin axis of B seems to be aligned with the orbital
angular momentum.

However, we have argued that the gravitomagnetic effect on time delay
still remains extremely small in this system: under reasonable assumptions
on the mass and angular momentum of the sources of the gravitational field,
the possibility to reveal the effect critically depends on the configuration of the
system and on the minimum impact parameter achievable for the e.m. ray.
Because of the existence of a large opaque region represented by a magnetosheath
surrounding PSR J0737-3039B, the effective impact parameter is much bigger than the
actual linear dimension of a neutron star, so that the magnitude of the gravitomagnetic
time delay is smaller than a reasonable detectability threshold.

\section*{ACKNOWLEDGMENTS}
The activity of A.N. in Val\`encia is supported by the
Angelo Della Riccia Foundation.


\end{document}